\documentclass[sigconf]{acmart}
\AtBeginDocument{%
  }

\setcopyright{None}

\acmConference[arXiv]{Preprint}{2026}{}

\usepackage{amsmath}
\usepackage{tikz}
\usetikzlibrary{matrix}
\usepackage{algorithm}
\usepackage{algpseudocode}
\usepackage{soul}
\usepackage{pgfplots}
\usepackage{booktabs}
\usepackage{graphicx} 

\newcommand{\best}[1]{\textbf{#1}}

\begin{document}
\sloppy
\title{You Don't Need Public Tests to Generate Correct Code}
\author{Kaushitha Silva}
\affiliation{%
  \institution{WSO2}
  \city{Santa Clara}
  \state{CA}
  \country{USA}
}
\email{kaushitha@wso2.com}
\orcid{0009-0005-0098-7815}

\author{Srinath Perera}
\affiliation{%
  \institution{WSO2}
  \city{Santa Clara}
  \state{CA}
  \country{USA}}
\email{srinath@wso2.com}
\orcid{0000-0002-4457-903X}


\begin{abstract}
    Multi-agent systems are frequently employed for autonomous code generation, demonstrating strong utility in complex algorithmic problem-solving. Recent studies tackle the difficulty of producing functionally correct programs by leveraging simulation-guided planning and debugging, wherein language models step through execution traces to validate logic. Nevertheless, these methods rely heavily on human-authored public test cases to anchor the simulation and debugging cycles. Hand-crafting exhaustive input-output pairs creates a significant, labor-intensive bottleneck within the software development lifecycle. Since ground-truth examples are seldom accessible before actual implementation in real-world scenarios, this reliance limits existing approaches primarily to curated competitive programming datasets. Additionally, we demonstrate that depending on these public tests creates an "overconfidence gap," leading frameworks to overfit to basic examples and underperform on hidden test suites. Conversely, we note that external input samples are not an absolute requirement for successful code generation. We show that large language models possess the capability to autonomously construct valid inputs and simulate execution flows for self-correction. Building on this, we introduce DryRUN, a framework that removes the necessity for ground-truth data by enabling the LLM to iteratively plan, synthesize its own test inputs, and run simulated executions, thereby mitigating algorithmic overconfidence. Assessments using the LiveCodeBench v6 dataset (post-March 2025) reveal that DryRUN achieves comparable performance to CodeSIM, a state-of-the-art, test-dependent baseline. Notably, it does so entirely without public tests or external execution signals, all while decreasing overall output token usage.
\end{abstract}

 



\begin{teaserfigure}
    \centering
    \begin{tikzpicture}[
        node distance=1.5cm and 2cm,
        spec/.style={rectangle, draw=black!40, rounded corners, fill=gray!5, text width=2.6cm, text centered, minimum height=0.9cm, font=\footnotesize\sffamily},
        abstract/.style={rectangle, draw=gray!60, dashed, rounded corners, fill=gray!5, text width=2.5cm, text centered, minimum height=0.9cm, font=\scriptsize\sffamily, text=gray!80!black},
        highlightA/.style={rectangle, draw=red!60!black, thick, rounded corners, fill=red!5, text width=3.4cm, text centered, minimum height=0.9cm, font=\footnotesize\bfseries\sffamily},
        highlightB/.style={rectangle, draw=blue!60!black, thick, rounded corners, fill=blue!5, text width=3.4cm, text centered, minimum height=0.9cm, font=\footnotesize\bfseries\sffamily},
        final/.style={rectangle, draw=green!50!black, rounded corners, fill=green!10, text width=1.8cm, text centered, minimum height=0.9cm, font=\footnotesize\sffamily},
        arrow/.style={->, thick, >=stealth, font=\scriptsize\sffamily, text=black!80}
    ]
    \node[font=\small\bfseries\sffamily, anchor=west] (titleA) at (-1.5, 1.3) {Standard Paradigm (e.g., CodeSIM)};
    \node[spec, draw=red!50!black, fill=red!5] (specA) at (0, 0) {Problem Spec + \textbf{Public Tests}};
    \node[abstract] (planA) at (3.8, 0) {Plan \& Generate Initial Code};
    \node[highlightA] (sandboxA) at (8.5, 0) {External Execution \textbf{Sandbox}};
    \node[final] (outA) at (13, 0) {Final Code};
    \node[highlightA, text width=4cm, minimum height=0.7cm] (traceA) at (6.15, -1.2) {Mental Trace on \textbf{Failed Public Test}};
    \node[font=\scriptsize\sffamily, text=red!70!black, align=center, text width=3.8cm] (riskA) at (13, -1.3) {\textbf{Higher Overconfidence Gap}};
    \draw[arrow] (specA) -- (planA);
    \draw[arrow] (planA) -- (sandboxA);
    \draw[arrow] (sandboxA) -- node[above, align=center, text width=2cm] {Passes Public Tests} (outA);
    \draw[arrow] (sandboxA.south) |- node[near start, right, align=center, text width=2cm] {Fails Public Tests} (traceA.east);
    \draw[arrow] (traceA.west) -| (planA.south);
    \draw[<-, dashed, red!70!black, thick, >=stealth] (riskA.north) -- (outA.south);
    \node[font=\small\bfseries\sffamily, anchor=west] (titleB) at (-1.5, -2.2) {DryRUN Paradigm (Ours)};
    \node[spec, draw=blue!50!black, fill=blue!5] (specB) at (0, -3.5) {Standardized Spec \textbf{(Zero Examples)}};
    \node[abstract] (planB) at (3.8, -3.5) {Plan \& Generate Initial Code};
    \node[highlightB] (synthB) at (8.5, -3.5) {\textbf{Autonomously} Synthesize Input};
    \node[final] (outB) at (13, -3.5) {Final Code};
    \node[highlightB, text width=4cm, minimum height=0.7cm] (traceB) at (6.15, -4.7) {Mental Trace \& Refine \textbf{(No Execution)}};
    \node[font=\scriptsize\sffamily, text=blue!70!black, align=center, text width=3.8cm] (genB) at (13, -4.8) {\textbf{Lower Overconfidence Gap}};
    \draw[arrow] (specB) -- (planB);
    \draw[arrow] (planB) -- node[above] {Start Loop} (synthB);
    \draw[arrow] (synthB.south) |- (traceB.east);
    \draw[arrow] (traceB.west) -| (planB.south);
    \draw[arrow] (planB.north) |- ++(0, 0.4) -| node[near start, above] {$N$ iterations completed} (outB.north);
    \draw[<-, dashed, blue!70!black, thick, >=stealth] (genB.north) -- (outB.south);
    \end{tikzpicture}
    \caption{State-of-the-art Code Generation methods (top) rely heavily on sample examples provided in the problem specification, passing code to an \textbf{external sandbox} and exclusively tracing code that \textbf{fails public tests}. As demonstrated later in our Overconfidence Gap analysis, this heavy reliance on trivial public tests often induces algorithmic overconfidence, leading to overfitting and subsequent failure on hidden private suites. In contrast, \textbf{DryRUN} (bottom) operates under strictly zero-example constraints. It shifts the burden of verification entirely inward, utilizing the LLM's ability to plan, autonomously synthesize inputs and simulate mental traces, eliminating the need for external oracles.}
    \Description{State-of-the-art Code Generation methods (top) rely heavily on sample examples provided in the problem specification, passing code to an external sandbox and exclusively tracing code that fails public tests. As demonstrated later in our Overconfidence Gap analysis, this heavy reliance on trivial public tests often induces overconfidence, leading to overfitting and subsequent failure on hidden private suites. In contrast, DryRUN (bottom) operates under strictly zero-example constraints. It shifts the burden of verification entirely inward, utilizing the LLM's ability to plan, autonomously synthesize inputs and execute mental traces, eliminating the need for external oracles.}
    \label{fig:teaser}
\end{teaserfigure}
\maketitle
\section{Introduction}

Program synthesis from natural language specifications represents a core objective in modern software engineering. Large language models (LLMs) have demonstrated remarkable proficiency in this domain, a capability primarily evaluated on standardized programming benchmarks such as HumanEval~\cite{chen_evaluating_2021} and LiveCodeBench~\cite{jain_livecodebench_2024}. However, the problem specifications within these benchmarks inherently embed human-authored sample input-output examples. Consequently, most code generation methods rely on these public samples for fundamental problem comprehension

Beyond this implicit reliance, recent state-of-the-art methods actively and explicitly exploit these public tests to drive their performance. This exploitation typically occurs across two distinct phases. During pre-generation reasoning, frameworks such as AlphaCodium~\cite{ridnik_code_2024} force the model to explicitly reason about how the provided public inputs map to their outputs to anchor the logic before writing any code. During post-generation debugging, methods utilize these public tests to obtain execution feedback. Furthermore, frameworks like CodeSIM~\cite{islam_codesim_2025} rely specifically on failed public test cases, prompting the model to mentally simulate an execution trace on that exact failed input to diagnose logical flaws. 

This deep dependency on public tests can create a massive disconnect from reality. In real-world software engineering contexts, ground-truth input-output examples are rarely available before the code is actually implemented. Furthermore, heavily relying on these tests introduces critical vulnerabilities. As observed by Liu et al.~\cite{liu_is_2023} and Liu et al. \cite{liu_rstar-coder_2025}, public test cases in standard benchmarks are often too trivial to catch complex edge cases. When models depend on signals from these simplistic tests, they frequently become overconfident in flawed solutions; passing the public tests but failing the hidden private suites. We formalize this disparity as the \textit{Overconfidence Gap}, which is illustrated in Figure 1 and analyzed extensively in Section~\ref{sec:discussion}. Conversely, manually authoring detailed public test cases that comprehensively cover edge cases is a tedious, labor-intensive process that places an unreasonable burden on developers.

\textit{These challenges force us to ask, are public test cases the only way?} Despite this ubiquitous implicit and explicit reliance on public tests, to the best of our knowledge, no prior studies have systematically evaluated how models perform when these sample examples are completely redacted from the prompt. 

Based on results we encountered while improving the planning for code generation, we hypothesize that this reliance on public tests is a crutch rather than a necessity. In this work, we first argue that removing human-authored examples from standard benchmarks yields negligible performance drops, indicating that modern LLMs do not strictly require them. Building on this insight, we introduce \textbf{DryRUN} (\textbf{D}ebugging and \textbf{R}efinement \textbf{U}nder \textbf{N}on-execution), a framework that completely replaces the need for public tests. DryRUN prompts the LLM to autonomously synthesize its own valid, hypothetical inputs and mentally simulate execution traces, allowing it to self-correct logical errors without relying on external examples or sandbox execution.

To rigorously evaluate our framework and mitigate the risk of data contamination, we conduct our experiments on the LiveCodeBench v6 dataset, strictly restricting our evaluation to problems published after March 2025. Because this benchmark natively includes public test cases, it provides an ideal testbed for a direct empirical comparison between standard public-test-dependent generation and our zero-example approach. We validate DryRUN's efficacy across two LLMs: \textit{gpt-5-mini}, \textit{gemini-3-flash}.

Our key contribution is questioning the role of public test cases as a major driver in correctness in code generation problems, exploring alternative approaches, and empirically demonstrating that the role of public test cases is much smaller than the community believes. 
Specifically, our main findings demonstrate that DryRUN matches the performance of CodeSIM, while operating under strictly zero-example constraints, no code execution, and consuming fewer tokens. Crucially, as depicted in Figure 1, DryRUN substantially mitigates the Overconfidence Gap by forcing the model to validate logic against its own non-trivial inputs rather than overfitting to trivial public oracles.
\begin{algorithm}[t]
\caption{The DryRUN Framework}
\label{alg:dryrun}
\renewcommand{\algorithmicrequire}{\textbf{Input:}}
\renewcommand{\algorithmicensure}{\textbf{Output:}}
\begin{algorithmic}[1]
\Require $\mathcal{P}$ (Standardized problem specification), $N_{plan}$, $N_{sim}$, $N_{refine}$, $\text{LLM}$
\Ensure $Code$

\State $Plan \gets \text{GeneratePlan}(\text{LLM}, \mathcal{P})$
\For{$i = 1$ \textbf{to} $N_{plan}$}
    \State $Plan \gets \text{RefinePlan}(\text{LLM}, \mathcal{P}, Plan)$
\EndFor

\State $Code \gets \text{SynthesizeCode}(\text{LLM}, \mathcal{P}, Plan)$

\For{$j = 1$ \textbf{to} $N_{sim}$}
    \State $Trace \gets \text{SimulateExecution}(\text{LLM}, \mathcal{P}, Code)$
    \State $Plan \gets \text{TraceDrivenRefine}(\text{LLM}, \mathcal{P}, Plan, Trace)$
    \State $Code \gets \text{SynthesizeCode}(\text{LLM}, \mathcal{P}, Plan)$
\EndFor
\State $Code \gets \text{PolishCode}(\text{LLM}, \mathcal{P}, Plan, Code)$
\State \Return $Code$
\end{algorithmic}
\end{algorithm}

\begin{figure*}[t]
    \centering
    \begin{tikzpicture}
        \matrix [
            matrix of nodes,
            inner sep=0pt,
            nodes={
                inner xsep=1.5ex,
                inner ysep=0.75ex,
                outer sep=0pt,
                align=left,
                font=\small\ttfamily,
                text width=0.92\textwidth
            },
            draw=black!30,
            thick
        ] (diff) {
            |[fill=black!5, text=black!60, font=\scriptsize\ttfamily]| \ \ @@ -1,14 +1,12 @@ \\
            \ \ You are given two positive integers, l and r. A positive integer is called beautiful if the product of its digits is divisible by the sum of its digits. \\
            \ \ Return the count of beautiful numbers between l and r, inclusive. \\
            |[fill=red!10, text=red!70!black]| - Example 1: \\
            |[fill=red!10, text=red!70!black]| - Input: l = 10, r = 20 \\
            |[fill=red!10, text=red!70!black]| - Output: 2 \\
            |[fill=red!10, text=red!70!black]| - Explanation: The beautiful numbers in the range are 10 and 20. \\
            |[fill=red!10, text=red!70!black]| - Example 2: \\
            |[fill=red!10, text=red!70!black]| - Input: l = 1, r = 15 \\
            |[fill=red!10, text=red!70!black]| - Output: 10 \\
            |[fill=red!10, text=red!70!black]| - Explanation: The beautiful numbers in the range are 1, 2, 3, 4, 5, 6, 7, 8, 9, and 10. \\
            |[fill=green!10, text=green!50!black]| + Input Format: \\
            |[fill=green!10, text=green!50!black]| +   - Two space-separated integers l and r. \\
            |[fill=green!10, text=green!50!black]| + Output Format: \\
            |[fill=green!10, text=green!50!black]| +   - A single integer: the count of beautiful numbers x such that l $\le$ x $\le$ r. \\
            \ \ Constraints: \\
            \ \ 1 $\le$ l $\le$ r < $10^9$ \\
        };
    \end{tikzpicture}
    \caption{LLM is prompted to remove the concrete sample examples (highlighted in red) and synthesize generalized input and output format descriptions (highlighted in green) to construct the zero-example problem specification $\mathcal{P}$.}
    \Description{LLM is prompted to remove the concrete sample examples (highlighted in red) and synthesize generalized input and output format descriptions (highlighted in green) to construct the zero-example problem specification $\mathcal{P}$.}
    \label{fig:problem_transformation}
\end{figure*}
\section{Related Work}
\subsection{Coding Benchmarks}

With the advent of Large Language Models (LLMs), the translation of natural language specifications into functional code has shown remarkable progress. Early evaluations relied heavily on static benchmarks such as HumanEval~\cite{chen_evaluating_2021}, MBPP~\cite{desai_program_2015}. Current contemporary models achieve pass@1 rates exceeding 90\% on these datasets. However, such metrics are increasingly suspect due to data contamination. To mitigate this, for a more accurate evaluation, recent benchmarks like LiveCodeBench~\cite{jain_livecodebench_2024}, LiveCodeBench-Pro~\cite{jain_livecodebench_2024}, SWE-Bench-Live~\cite{zhang_swe-bench_2025}, and LiveBench~\cite{white_livebench_2025} evaluate models exclusively on problems released after their respective training cutoffs. Experiments reveal that state-of-the-art models exhibit substantial performance degradation on these uncontaminated datasets, indicating that zero-shot inference remains insufficient to guarantee correctness on novel tasks. A key aspect of these coding benchmarks is that they provide a small set of ground-truth sample input-output pairs to clarify the problem scope and expected formatting. In LiveCodeBench, these examples appear both in the problem specification, often with some explanation, and as public test cases which can be used for test execution against the synthesized code. 

\subsection{LLM-Driven Code Generation Techniques}

Despite the remarkable zero-shot capabilities of LLMs, direct generation often yields subtle syntactic or logical flaws. Initial efforts to rectify these issues focused solely on prompting strategies, such as Chain-of-Thought (CoT)~\cite{wei_chain--thought_2023} Structured CoT (SCoT)~\cite{li_structured_2023}, and self-planning~\cite{jiang_self-planning_2024}, which demonstrated that intermediate reasoning improves generation quality. While these methods improved performance, they did not explicitly utilize execution environments or sample examples. 

Consequently, closed-loop methods emerged, utilizing execution feedback to iteratively debug generated code. Early self-debugging approaches extracted the few provided public sample examples to execute and refine the code. Recognizing that two or three examples are rarely comprehensive, subsequent frameworks like AgentCoder~\cite{huang_agentcoder_2024}, Reflexion~\cite{shinn_reflexion_2023}, ThinkCoder~\cite{zhang_thinking_2025}, DebateCoder~\cite{chen_debatecoder_2025}, CodeCoR~\cite{pan_codecor_2025}, and CodeT~\cite{chen_codet_2022} prompted LLMs to generate extensive autonomous test suites. CodeT, for example, uses a dual-agreement between generated test cases and code to select a solution out of many. Similarly, S*~\cite{li_s_2025} generates several candidate solutions and compares them as pairs, using differential inputs to select the best candidate based on execution behavior. 

However, as noted by the authors of AgentCoder themselves, these LLM-generated test cases are frequently flawed, leading to false-positive acceptance or misdirected repairing of the generated code. To address this unreliability, frameworks like AlphaCodium~\cite{ridnik_code_2024} introduced an anchoring mechanism, explicitly grounding the generated test cases by utilizing the human-authored sample input-output pairs.

\subsection{Dependency on Public Test Cases}

To circumvent the inherent unreliability of LLM-generated tests, the most recent generation of state-of-the-art frameworks, such as MapCoder~\cite{islam_mapcoder_2024} and CodeSIM~\cite{islam_codesim_2025}, has largely abandoned autonomous test generation. Instead, they rely almost exclusively on the sample I/O (public test cases) provided within the problem specification. CodeSIM, in particular, simulates the execution trace of the generated program, specifically when it fails these public tests, grounding the LLM's debugging logic in its own mental simulation of the failure. Once the public test cases pass, these frameworks accept the code as the final solution. Furthermore, an ablation study within CodeSIM demonstrates that the performance increases with the number of public test cases. 

However, as discussed in the introduction, this heavy reliance on public tests introduces critical vulnerabilities. As observed by Liu et al.[1], the public test cases in standard benchmarks are often trivial and fail to cover complex edge cases. When models depend on signals from these simplistic tests, they frequently become overconfident in flawed solutions. Furthermore, creating detailed public test cases that adequately capture edge cases is a tedious, labor-intensive process that places an unreasonable burden on developers. Most importantly, relying on these examples creates a stark disconnect from real-world software engineering, where comprehensive input-output examples are rarely available prior to implementation.

DryRUN explores an alternative to this critical dependency by leveraging the internal simulation capabilities of modern Large Language Models (LLMs). Existing methods typically require either unverified generated tests or human-provided public tests to guide refinement. In contrast, we investigate whether this requirement can be relaxed. By prompting the model to synthesize hypothetical inputs and simulate corresponding execution traces, DryRUN enables trace-driven refinement in a zero-example setting.

\section{The DryRUN Framework}

The DryRUN framework operates through a multi-phase loop comprising initial planning, iterative plan refinement, code synthesis, trace-driven refinement via mental simulation, and final polishing, as depicted by Algorithm~\ref{alg:dryrun}.

\subsection{Phase 1: Initial Implementation Planning}
Drawing upon established literature, we initially prompt the LLM to formulate a step-by-step implementation plan derived directly from the problem specification $\mathcal{P}$. Rather than immediately generating code, we subject this initial plan to $N_{plan}$ iterations of autonomous refinement without any external feedback. As demonstrated in our ablation studies (Section~\ref{subsec:ablation}) this isolated plan refinement loop systematically eliminates preliminary logical oversights and independently improves overall code generation performance.

\subsection{Phase 2: Mental Simulation and Trace-Driven Refinement}
After synthesizing an initial code draft from the refined plan, the core simulation loop executes for $N_{sim}$ iterations. In each iteration, the LLM performs a mental dry run by autonomously synthesizing a valid, non-trivial sample input based purely on the problem constraints defined in $\mathcal{P}$. The model subsequently simulates the execution of the current code line-by-line against this generated input, tracking variable states to expose hidden logical flaws. The resulting execution trace is fed back into the planning module, which explicitly updates the implementation plan to resolve the identified bugs. Finally, the code is regenerated to reflect these trace-driven corrections. Crucially, in contrast to recent trace-based methods such as CodeSIM, DryRUN provides neither human-authored sample I/O nor sandbox execution feedback; the LLM relies entirely on its intrinsic capacity to formulate inputs and mentally simulate their execution paths.

\subsubsection{Phase 3: Final Code Polishing}
To conclude the generation process, the framework executes a final polishing stage. The LLM reviews the finalized plan alongside the latest code iteration to resolve any syntax anomalies, improve stylistic clarity, and ensure absolute alignment with the plan. This phase produces the final, self-corrected solution. As detailed in our ablation studies (Section~\ref{subsec:ablation}), this final pass on the  code serves as an effective mechanism for rectifying any errors that may have been introduced while correcting the code.

\subsection{DryRUN vs CodeSIM}

Although both DryRUN and CodeSIM follow a similar simulation phase, there are key fundamental differences that separates the two approaches. 
\begin{itemize}
    \item \textbf{Autonomous Plan Refinement: } DryRUN consists of a plan refinement phase with no external feedback. The plan is refined by prompting the LLM to improve the logic, add more details steps and consider more edge cases. 
    \item \textbf{Execution-Free Refinement: } DryRUN strictly omits external code execution, the generated code is never executed against any input. 
    \item \textbf{Self-generated Inputs: } DryRUN does the mental simulation of a non-trivial input that the LLM comes up with its own, whereas in CodeSIM, the input is chosen from the failed public test cases. 
    \item \textbf{Simulation is preceded by plan refinement: } DryRUN always refines the plan and then generates the code. In comparison CodeSIM prompts the LLM to either repair the code or the plan. 
    \item \textbf{Final Code Refinement: } DryRUN finally polishes the refined code, this is to compensate for the lack of execution where there may have been syntax or logic errors introduced in the refining stage. 
\end{itemize}
\section{Experiments and Results}
\begin{table*}[t]
  \centering
  \small
  \caption{Comparison of Pass@1 (\%) performance on LiveCodeBench V6 (Post-March 2025) averaged across 3 independent runs. Values are reported as mean $\pm$ standard deviation.}
  \label{tab:main_results}
  \setlength{\tabcolsep}{5pt}
  \begin{tabular}{l cccc c cccc}
    \toprule
    & \multicolumn{4}{c}{\textbf{gpt-5-mini}} & & \multicolumn{4}{c}{\textbf{gemini-3-flash}} \\
    \cmidrule(lr){2-5} \cmidrule(lr){7-10}
    \textbf{Method} & Easy & Med & Hard & Overall & & Easy & Med & Hard & Overall \\
    \midrule
    Direct (w/ Public)  & 83.3 $\pm$ 5.6 & 40.0 $\pm$ 4.0 & 20.7 $\pm$ 4.1 & 40.8 $\pm$ 1.9 & &  92.6 $\pm$ 3.2 & 65.3 $\pm$ 2.3 & 41.4 $\pm$ 4.1 & 60.4 $\pm$ 1.9 \\
    Direct (w/o Public)  & 75.9 $\pm$ 3.2 & 52.0 $\pm$ 10.6& 27.9 $\pm$ 3.1 & 46.3 $\pm$ 4.5 & &  98.2 $\pm$ 3.2 & 69.3 $\pm$ 2.3 & 41.4 $\pm$ 3.1 & 62.9 $\pm$ 0.7 \\
    CodeSIM          & \best{92.6} $\pm$ 3.2 & \best{77.3} $\pm$ 8.3 & 41.4 $\pm$ 9.5 & 64.2 $\pm$ 7.5 & & \best{100.0} $\pm$ 0.0 & \best{81.3} $\pm$ 2.3 & \best{57.7} $\pm$ 4.1 & \best{74.6} $\pm$ 1.9 \\
    \midrule
    \textbf{DryRUN (Ours)} & 90.7 $\pm$ 3.2 & 72.0 $\pm$ 6.9 & \best{53.2} $\pm$ 5.6 & \best{67.5} $\pm$ 2.5 & & \best{100.0} $\pm$ 0.0 & 77.3 $\pm$ 8.3 & 49.6 $\pm$ 1.6 & 69.6 $\pm$ 3.1 \\
    \bottomrule
  \end{tabular}
\end{table*}
\begin{table}[htbp]
  \centering
  \small
  \caption{Comparison of average token consumption per problem. Values represent the mean token counts consumed.}
  \label{tab:token_costs}
  \setlength{\tabcolsep}{8pt}
  \begin{tabular}{l rrr}
    \toprule
    \textbf{Method} & \textbf{Input} & \textbf{Output} & \textbf{Total} \\
    \midrule
    Direct & 573 & 896 & 1469 \\
    CodeSIM            & 18,922 & 22,054 &	40,976 \\
    \midrule
    \textbf{DryRUN (Ours)} & 17,897 & 10,670 & 28,567 \\
    \bottomrule
  \end{tabular}
\end{table}
\subsection{Experimental Setup}
\subsubsection{Models}
We evaluate our framework using two recent large language models: \textit{gpt-5-mini} and \textit{gemini-3-flash}, both configured to use ``minimal'' reasoning effort. For \textit{gemini-3-flash}, we set the temperature to its default value of 1.0. However, the temperature of the gpt-5-mini model is not configurable, hence it is left to its default. 

We also conducted experiments with an open-weight Small Language Model (SLM), \textit{qwen2.5-coder:7b}. DryRUN failed many problems, passing none of the hard problems. Analyzing the SLM responses revealed that it was unable to follow complex prompts, and the generated code contained frequent syntax anomalies. Hence, we did not conduct further tests on SLMs.

\subsubsection{Dataset Selection}
To accurately evaluate code generation, we needed a contamination-free benchmark that is published after the model's data cutoff. Furthermore, to test the specific impact of public test cases, the dataset must include human-authored sample inputs and outputs. Based on these criteria, we select the LiveCodeBench-v6~\cite{jain_livecodebench_2024} dataset, strictly limiting our evaluation to problems published after March 2025. This cutoff ensures a contamination-free evaluation, as the most recent training cutoff among our evaluated models (\textit{gemini-3-flash}) is reported as January 2025. This filtering yields a final set of 80 problems, consisting of 37 hard, 25 medium, and 18 easy problems. We exclude other recent live benchmarks, such as SWELive~\cite{zhang_swe-bench_2025}, because they do not provide human-authored sample input-output pairs.

\subsubsection{Removal of Examples of the Problem Specification}

Problem descriptions in code generation benchmarks typically include input-output examples and sometimes reasoning steps. We prompt an LLM to replace these elements with descriptions of the expected input and output formats. We manually verified the dataset to ensure no examples or ground-truth data remain. This process yields the standardized, example-free problem specification $\mathcal{P}$ used throughout our experiments (see Figure~\ref{fig:problem_transformation} for a representative example).

\subsubsection{DryRUN Configuration}
For our primary results, we configure the DryRUN framework with $N_{plan} = 2$ planning iterations and $N_{sim} = 2$ simulation iterations. All ablation studies are conducted exclusively using \textit{gpt-5-mini}.

\subsubsection{Baselines}
We first compare DryRUN against standard direct (zero-shot) generation in two settings: with and without the inclusion of public test cases in the prompt. Additionally, we benchmark against CodeSIM, the current state-of-the-art multi-agent framework that utilizes execution simulation as per Islam et al.~\cite{islam_codesim_2025}. We execute three independent runs for each problem and report the average pass@1 success rate.

\subsection{Results}

As indicated in Table~\ref{tab:main_results}, DryRUN and CodeSIM achieve comparable overall performance. DryRUN outperforms CodeSIM by $3.3\%$ utilizing \textit{gpt-5-mini}, whereas CodeSIM exhibits a $5.0\%$ advantage with \textit{gemini-3-flash}, with both performance margins falling within overlapping standard deviations. However, since DryRUN operates without public test cases, this performance parity strongly supports our core hypothesis: planning and simulation can effectively compensate for the absence of public tests. This empirical evidence suggests that the fundamental reliance on human-authored public tests is much smaller than conventionally assumed by the community. Furthermore, excising public tests from the zero-shot baseline (Direct w/o Public) slightly improved baseline accuracy, further reinforcing that models do not strictly require these examples for task comprehension.

As expected, the pass@1 performance decreases with problem difficulty, accompanied by an increase in variance. This increase in variance is likely attributable to the increase in complexity of execution simulation on harder problems, where the impact of the model's probabilistic nature during mental tracing becomes more pronounced. Across all configurations, \textit{gemini-3-flash} consistently demonstrated superior code generation capabilities compared to \textit{gpt-5-mini}.

Regarding computational overhead indicated in Table~\ref{tab:token_costs}, both simulation-driven frameworks consume more compute than zero-shot generation. While DryRUN and CodeSIM are approximately $19\times$ and $27\times$ more token-intensive than the Direct baseline, respectively, DryRUN is notably more efficient. DryRUN consumes only $70\%$ of CodeSIM's total tokens, and merely $50\%$ of the output tokens, which are typically the primary bottleneck for inference latency and cost. A likely explanation for this is that CodeSIM's default configuration permits up to 5 simulations for planning and an additional 5 for code repair, meaning a difficult problem could trigger up to $25$ nested simulations. In contrast, DryRUN strictly enforces a maximum of two simulation rounds ($N_{sim}=2$).

Overall, these results demonstrate that even without public test cases, DryRUN achieves competitive performance while maintaining a consuming fewer tokens.

\subsection{Ablation Studies}
\label{subsec:ablation}

To systematically isolate the contribution of each algorithmic component within the DryRUN framework, we conduct a rigorous ablation study. The first ablation switches from zero-shot generation to a structured base planning, while the second and third add planning refinement steps. Fourth and fifth ablations add mental simulation. The final ablation is the Full DryRUN configuration including the final code polishing stage. Table~\ref{tab:ablation_results} depicts the results.

All ablations are performed using the \textit{gpt-5-mini} model evaluated exclusively on the Hard subset of LiveCodeBench V6 ($N=37$). To account for variance, we report the mean Pass@1 rate and standard deviation across three independent runs. 

\begin{table}[htbp]
  \centering
  \small
  \caption{Ablation study of the DryRUN framework on the LiveCodeBench Hard subset using gpt-5-mini. Results are averaged across 3 runs. The nomenclature denotes the number of plan refinement iterations ($N_{plan}$) and mental simulation iterations ($N_{sim}$).}
  \label{tab:ablation_results}
  \setlength{\tabcolsep}{8pt}
  \begin{tabular}{l c}
    \toprule
    \textbf{Framework Configuration} & \textbf{Pass@1 (\%) $\pm$ SD} \\
    \midrule
    Direct (Zero-Shot Baseline) & 27.93 $\pm$ 3.12 \\
    \midrule
    Base Planning & 43.24 $\pm$ 4.68 \\
    + Plan Refinement ($N_{plan}=1$) & 45.95 $\pm$ 5.41 \\
    + Plan Refinement ($N_{plan}=2$) & 48.65 $\pm$ 2.70 \\
    \midrule
    + Mental Simulation ($N_{plan}=2$, ${N_{sim}=1}$) & 47.75 $\pm$ 1.56 \\
    + Mental Simulation ($N_{plan}=2$, ${N_{sim}=2}$) & 49.55 $\pm$ 1.56 \\
    \midrule
    \textbf{+ Final Polish (Full DryRUN)} & \best{53.15} $\pm$ \best{5.63} \\
    \bottomrule
  \end{tabular}
\end{table}

\subsection{The Impact of Isolated Planning}
The transition from direct, zero-shot generation to a structured base planning phase yields a baseline improvement of $+15.31\%$. This aligns with established literature indicating that large language models require intermediate reasoning steps to effectively manage the logic of complex algorithms~\cite{wei_chain--thought_2023, huang_codecot_2024, jiang_self-planning_2024, li_structured_2023}. Iterating on this plan without external feedback demonstrates consistent empirical gains. Advancing from a single planning step to two refinement iterations ($N_{plan}=2$) increases the Pass@1 rate to $48.65\%$ while halving the standard deviation (from $5.41$ to $2.70$). This suggests that isolated plan refinement likely helps stabilize the model's logic and mitigate preliminary algorithmic oversights prior to code synthesis.

\subsection{Mental Simulation}
A single iteration of mental simulation ($N_{plan}=2$, $N_{sim}=1$) results in a marginal regression in mean accuracy compared to the purely planning-based approach. This likely stems from the inherent difficulty of autonomous patching: when the model attempts to correct complex logic based on its initial mental trace, it can occasionally introduce syntactic or logical regressions. However, allowing the simulation loop to execute for a second iteration ($N_{plan}=2$, $N_{sim}=2$) helps the model stabilize these patches. This recovers the mean performance to 49.55\% while critically suppressing the variance to a highly consistent ± 1.56. This leads us to hypothesize that while the first simulation iteration helps expose hidden flaws, a subsequent iteration provides a valuable verification step to stabilize the applied fixes.

\subsection{The Necessity of Compensatory Polishing}
The addition of the final code polishing stage triggers a $+3.60\%$ performance spike, achieving the peak Pass@1 rate of $53.15\%$. We hypothesize that, while the mental simulation phase is highly effective at identifying logical edge cases, the process of explicitly updating the plan and regenerating code without a live compiler can introduce syntactic and logical anomalies. It appears the final polishing step acts as a compensatory analytical pass, helping to smooth out these minor syntactic and logical deviations. 

\section{Discussion}
\label{sec:discussion}
\begin{figure}[htbp]
    \centering
    \begin{tikzpicture}
        \begin{axis}[
            ybar stacked,
            bar width=16pt,
            width=8.5cm, 
            height=7.5cm,
            enlarge x limits=0.2,
            ymin=0, ymax=40,
            ylabel={Number of Problems},
            xtick={1, 2, 3, 4},
            xticklabels={
                Direct\\(Public), 
                Direct\\(No Pub), 
                CodeSIM, 
                \textbf{DryRUN\\(Ours)}
            },
            xticklabel style={align=center, text width=1.8cm, font=\small, yshift=-1ex},
            legend style={
                at={(0.02,0.98)}, 
                anchor=north west, 
                legend columns=1, 
                draw=gray!50, 
                fill=white, 
                font=\scriptsize
            },
            major x tick style={draw=none},
            axis x line*=bottom,
            axis y line*=left,
            ymajorgrids=true,
            grid style={dashed, gray!30},
            clip=false 
        ]

        \addplot[
            fill=gray!90!black, 
            draw=black!80,
            error bars/.cd,
                y dir=both, 
                y explicit,
                error bar style={thick, black, xshift=0pt},
                error mark options={rotate=90, mark size=3pt, thick}
        ] coordinates {
            (1, 7.67)  +- (0, 1.53)  
            (2, 10.33) +- (0, 1.15)  
            (3, 15.33) +- (0, 3.51)  
            (4, 19.67) +- (0, 2.08)  
        };
        
        \addplot[
            fill=red!60!white, 
            draw=black!80,
            error bars/.cd,
                y dir=both, 
                y explicit,
                error bar style={thick, black, xshift=0pt},
                error mark options={rotate=90, mark size=3pt, thick}
        ] coordinates {
            (1, 7.00)  +- (0, 0.00)  
            (2, 7.67)  +- (0, 2.08)  
            (3, 14.00) +- (0, 4.36)  
            (4, 7.00)  +- (0, 0.00)  
        };

        \legend{Actually Solved (Private Passed), Overconfidence Gap (Public Only)}
        \end{axis}
    \end{tikzpicture}
    \caption{ Analysis of the Overconfidence Gap (gpt-5-mini, Hard Subset, $N=37$). The red stack indicates problems that pass public tests but fail the hidden private suite. CodeSIM displays a large, highly volatile gap ($\pm 4.36$), indicating overfitting to trivial oracles.}
    \Description{Analysis of the Overconfidence Gap (gpt-5-mini, Hard Subset, $N=37$). The red stack (with standard deviation error bars across 3 runs) indicates problems that pass public tests but fail the hidden private suite. CodeSIM displays a large, highly volatile gap ($\pm 4.36$), indicating overfitting to trivial oracles.}
    \label{fig:single_col_overconfidence}
\end{figure}
\begin{figure}[htbp]
    \centering
    \begin{tikzpicture}
        \begin{axis}[
            ybar stacked,
            bar width=16pt,
            width=8.5cm, 
            height=7.5cm,
            enlarge x limits=0.2,
            ymin=0, ymax=40,
            ylabel={Number of Problems},
            xtick={1, 2, 3, 4},
            xticklabels={
                Direct\\(Public), 
                Direct\\(No Pub), 
                CodeSIM, 
                \textbf{DryRUN\\(Ours)}
            },
            xticklabel style={align=center, text width=1.8cm, font=\small, yshift=-1ex},
            legend style={
                at={(0.02,0.98)}, 
                anchor=north west, 
                legend columns=1, 
                draw=gray!50, 
                fill=white, 
                font=\scriptsize
            },
            major x tick style={draw=none},
            axis x line*=bottom,
            axis y line*=left,
            ymajorgrids=true,
            grid style={dashed, gray!30},
            clip=false 
        ]

        \addplot[
            fill=gray!90!black, 
            draw=black!80,
            error bars/.cd,
                y dir=both, 
                y explicit,
                error bar style={thick, black, xshift=0pt},
                error mark options={rotate=90, mark size=3pt, thick}
        ] coordinates {
            (1, 15.33) +- (0, 1.53)  
            (2, 15.33) +- (0, 1.15)  
            (3, 21.33) +- (0, 1.53)  
            (4, 18.33) +- (0, 0.58)  
        };
        
        \addplot[
            fill=red!60!white, 
            draw=black!80,
            error bars/.cd,
                y dir=both, 
                y explicit,
                error bar style={thick, black, xshift=0pt},
                error mark options={rotate=90, mark size=3pt, thick}
        ] coordinates {
            (1, 6.33)  +- (0, 2.08)  
            (2, 7.67)  +- (0, 1.15)  
            (3, 12.00) +- (0, 1.73)  
            (4, 6.67)  +- (0, 2.08)  
        };

        \legend{Actually Solved (Private Passed), Overconfidence Gap (Failed Private)}
        \end{axis}
    \end{tikzpicture}
    \caption{Analysis of the Overconfidence Gap (gemini-3-flash, Hard Subset, $N=37$). The red stack indicates problems that pass public tests but fail the hidden private suite. Similar to the \textit{gpt-5-mini} results, CodeSIM displays an inflated overconfidence gap, whereas DryRUN maintains a lower gap ($\pm 2.08$), even though the pass@1 performance is lower.}
    \label{fig:single_col_overconfidence_gemini}
    \Description{
    Analysis of the Overconfidence Gap (gemini-3-flash, Hard Subset, $N=37$). The red stack indicates problems that pass public tests but fail the hidden private suite. Similar to the \textit{gpt-5-mini} results, CodeSIM displays an inflated overconfidence gap, whereas DryRUN maintains a lower gap ($\pm 2.08$), even though the pass@1 performance is lower.
    }
\end{figure}

\subsection{What could explain the results?}

To understand how DryRUN matches CodeSIM's performance without relying on public test cases, we must contrast their underlying refinement mechanisms. During the initial planning phase, CodeSIM conditionally validates its plan via simulation, refining it only if the simulation exposes a logical flaw. In contrast, DryRUN employs a proactive approach, unconditionally refining its plan $N_{plan}=2$ times without any simulation. During the code generation phase, CodeSIM relies on a reactive, external execution based workflow: it executes the generated code against the public test suite and accepts the solution if the public tests pass. If they fail, it mentally traces a failing public test to attempt a repair. Conversely, DryRUN never executes the code against an external sandbox. Instead, it unconditionally performs $N_{sim}=2$ rounds of mental simulation using autonomously synthesized inputs, continuously updating the plan and regenerating the code regardless of successful or failed simulation. 

This divergence leads to distinct outcomes. If the provided public test cases are trivial or fail to cover complex edge cases, CodeSIM is adversely affected; because it relies exclusively on these public tests for validation, it risks becoming overconfident in flawed solutions. Instances where the generated code passes the public tests but fails the hidden private suite are indicative of this behavior. We term this phenomenon as the \textbf{Overconfidence Gap}. As shown in Figures~\ref{fig:single_col_overconfidence} and \ref{fig:single_col_overconfidence_gemini}, which illustrate this gap for Hard problems, CodeSIM exhibits a larger overconfidence gap than all other methods (including DryRUN) across both LLMs. Therefore, we believe this algorithmic overconfidence is a plausible explanation for the performance differences. In contrast, DryRUN maintains a distinct advantage by unconditionally simulating the code against its own synthesized inputs, thereby mitigating the risk of overfitting.

On the other hand, when public test cases contain complex edge cases, CodeSIM's reactive simulation provides a distinct advantage. If the generated code passes these tests, it is highly likely that the correct solution is achieved. In failure cases, CodeSIM will reactively try to simulate these edge case inputs. Again, if it is capable of simulating and mentally tracing the edge case, it is likely to produce the correct solution. However, we noticed two key observations which suggested otherwise. 
\begin{itemize}
    \item When the input is complex, the LLM struggles to mentally trace the output. This was commonly seen in math heavy problems. (e.g: \texttt{arc195\_e}: algorithm contains a large modular division) and with larger inputs (e.g: \texttt{arc194\_c} - 3rd public test case)
    \item Since the prompt contains the error log from the public test execution, the LLM already knows the output. We noticed cases where it did not simulate but simply concluded that the code was incorrect, instead of tracing step-by-step as instructed. (e.g: \texttt{arc195\_e}: see Figure~\ref{fig:sim_abandonment})
\end{itemize}

\begin{figure}[htbp]
    \centering
    \begin{tikzpicture}
        \node[
            rectangle, 
            draw=gray!50, 
            fill=gray!5, 
            rounded corners=4pt, 
            inner sep=8pt, 
            text width=0.95\columnwidth, 
            font=\scriptsize\ttfamily
        ] (log) {
\textcolor{gray!80!black}{[... CodeSIM Simulation Excerpt ...]}\\[1ex]
Take the provided failing test case input:\\
9 6\\
765689282 93267307 563699854 951829154\\
801512848 389123318 924504746 596035433\\
3 8\\
2 5\\
5 8\\
2 9\\
8 9\\
5 7\\[1.5ex]
Run through the code step by step:\\[1ex]
1. Parse N=9, Q=6.\\
2. Read A[2..9] modulo MOD:\\
\hphantom{2. }A[2]=765689282, A[3]=93267307,\\
\hphantom{2. }A[4]=563699854, A[5]=951829154,\\
\hphantom{2. }A[6]=801512848, A[7]=389123318,\\
\hphantom{2. }A[8]=924504746, A[9]=596035433.\\
3. Compute fact = (N-1)! = 8! modulo MOD.\\
4. Precompute inv[i] = pow(i, MOD-2, MOD) for i=1..N.\\
5. Build prefix S[i] = sum\_\{j=2..i\} A\_j * inv[j-1] (mod MOD).\\
6. For each query (u,v) compute val = (S[v] - S[u]) \% MOD and ans = fact * val \% MOD.\\[1.5ex]
\textcolor{red!70!black}{When running this, the produced outputs differ from the expected outputs (they were reported as failing). So the code produced incorrect answers for that test case.}\\[1ex]
\textcolor{gray!80!black}{[... Log Truncated ...]}
        };
    \end{tikzpicture}
    \caption{\textbf{Example of a failure case of CodeSIM's Simulation. } An excerpt from a generated simulation log on a complex, math-heavy problem (\texttt{arc195\_e}). When confronted with large integer arrays and modular arithmetic, the LLM successfully outlines the high-level algorithmic steps but fails to perform the step-by-step trace and reach an output. Biased by the explicit error signal provided in the prompt, it halts the simulation (highlighted in red) and concludes the code is incorrect without pinpointing the actual logical flaw.}
    \label{fig:sim_abandonment}
    \Description{
    Example of a failure case of CodeSIM's Simulation. An excerpt from a generated simulation log on a complex, math-heavy problem (arc195\_e). When confronted with large integer arrays and modular arithmetic, the LLM successfully outlines the high-level algorithmic steps but fails to perform the step-by-step trace and reach an output. Biased by the explicit error signal provided in the prompt, it halts the simulation (highlighted in red) and concludes the code is incorrect without pinpointing the actual logical flaw.
    }
\end{figure}

\begin{figure}[t]
    \centering
    \begin{tikzpicture}
        \node[
            rectangle, 
            draw=gray!50, 
            fill=gray!5, 
            rounded corners=4pt, 
            inner sep=8pt, 
            text width=0.95\columnwidth, 
            font=\scriptsize\ttfamily
        ] (log) {
\textcolor{gray!80!black}{[... DryRUN Simulation Excerpt ...]}\\[1ex]
\textcolor{blue!70!black}{1) Choose a concrete sample input}\\[1ex]
We must obey constraints (2 $\le$ N $\le$ 2e5, Q up to 2e5).\\
To be non-trivial but small enough to trace manually, pick:\\
- N = 5\\
- Q = 4\\
- A = (A2..A5) = [2, 3, 5, 7]\\
- Queries: (1,2), (1,5), (2,4), (3,5)\\[1.5ex]
So input (as lines) is:\\
5 4\\
2 3 5 7\\
1 2\\
1 5\\
2 4\\
3 5\\[1.5ex]
\textcolor{blue!70!black}{2) Parse input and initial setup}\\[1ex]
data sequence:\\
{[}5, 4, 2, 3, 5, 7, 1, 2, 1, 5, 2, 4, 3, 5{]}\\[1ex]
N = next(it) = 5\\
Q = next(it) = 4\\[1.5ex]
\textcolor{gray!80!black}{[... Step-by-step arithmetic trace ...]}\\[1.5ex]
Query4: (3,5)\\
u=3, v=5\\
part1: x in 2..u-1 => 2..2 => S1 = B1[2]=24; S12 = B12[2]=8; \\
\hphantom{part1: }diff=16; part1 = 2*16=32\\
part2: BFT[3] - B1[3] = 72 - 24 = 48\\
part3: x in 4..4 => B1[4] = 30\\
part4: BFT[5] = 168\\[1ex]
ans = 32 + 48 + 30 + 168 = 278\\[1.5ex]
So outputs lines:\\
48\\
246\\
168\\
278\\[1ex]
\textcolor{gray!80!black}{[... Log Truncated ...]}
        };
    \end{tikzpicture}
    \caption{\textbf{Example of Simulation of DryRUN} An excerpt from DryRUN's simulation for the same problem (\texttt{arc195\_e}). The LLM generates a non-trivial but traceable input. This self-calibration ensures the subsequent step-by-step arithmetic trace remains well within its intrinsic simulation capacity, allowing it to successfully compute the final outputs.}
    \label{fig:input_synthesis}
    \Description{
    An excerpt from DryRUN's simulation for the same problem (\texttt{arc195\_e}). The LLM generates a non-trivial but traceable input. This self-calibration ensures the subsequent step-by-step arithmetic trace remains well within its intrinsic simulation capacity, allowing it to successfully compute the final outputs.
    }
\end{figure}

In contrast as shown in Figure~\ref{fig:input_synthesis}, DryRUN forces the LLM to autonomously synthesize its own non-trivial inputs. We observed a crucial phenomenon, while the model generates valid edge cases, it inherently avoids synthesizing inputs that are too massive or complicated for it to mentally trace. We hypothesize that the LLM naturally calibrates the complexity of the generated inputs to strictly align with its intrinsic simulation capacity. This self-regulating behavior ensures that DryRUN consistently executes stable, productive mental traces without becoming overwhelmed, further explaining its competitive performance.

We hypothesize that these contrasting advantages: DryRUN's proactive refinements based on non-trivial but simpler inputs  versus CodeSIM's ability to leverage public tests with edge cases, counterbalance one another in our experiments, resulting in the observed performance parity. We intend to explore this dynamic further in future work.


\subsection{DryRUN and SLMs}
DryRUN did not do well with SLMs. With (\textit{qwen2.5-coder:7b}), it only passed 6/18 of the easy problems and was far inferior to CodeSIM, which passed 13/18 problems. Inspecting the execution traces showed that the model frequently produced syntax errors and did not follow complex multi-step instructions (e.g., although instructions ask only to plan, not to create code, it often produced code in the first step. So they often failed one of many complex steps. This is likely because of the gap between the reasoning capabilities of SLMs and LLMs. Recent work has pointed out that SLMs in focused domains can reason much better than their size suggests~\cite{wang_comprehensive_2024}. As future work, we would like to explore how to tweak SLMs to provide good results with DryRUN.  

\subsection{Should we drop public tests from benchmarks?}
Furthermore, these results bring into question whether we should include public test cases in the benchmarks. As we discussed, providing public test cases is tedious, and weak public test cases may lead to overconfidence in code generation frameworks, leading to failures. The availability of public test cases will bias code-generation solutions toward their availability. However, in real world scenario, it is likely that many developers will only provide trivial public examples, which can lead to reduced performance in the field. While we do not believe our results are sufficient to conclude that we should remove public test cases from the benchmarks, it raises the question.

\subsection{Future Work}

Future research encompasses three primary directions. First, adapting mental trace-driven refinement for Small Language Models (SLMs) remains an open challenge. Initial experiments indicate that current SLMs struggle with the multi-step reasoning required for mental simulation, and inability to follow complex commands (e.g: writing code when simulating, writing plans when prompted to refine the code), and frequently introducing syntax errors when generating the code. Second, while DryRUN strictly omits external execution, the framework inherently yields autonomously synthesized inputs and mentally predicted outputs. Reintroducing an execution sandbox purely to verify the accuracy of these mental simulations presents a promising diagnostic mechanism. Discrepancies between the model's mental trace and the deterministic sandbox execution would yield a high-fidelity feedback signal, explicitly highlighting flaws in the model's comprehension of the problem specification. Finally, scaling the DryRUN framework to repository-level tasks (e.g., SWE-bench~\cite{jimenez_swe-bench_2024}) represents a necessary evolution. Because real-world repositories inherently lack comprehensive input-output oracles, applying mental simulation to resolve cross-file dependencies serves as a rigorous testbed for the practical applicability of zero-example code generation. 
\section{Conclusion}

Current code generation frameworks heavily depend on human-provided public tests, creating a disconnect from real-world software engineering where comprehensive input-output examples are often unavailable. We argued that relying on these simplistic tests frequently can cause the model to become overconfident in flawed solutions that fail to cover complex edge cases. To address this, DryRUN operates under strictly zero-example constraints by iteratively planning, autonomously synthesizing valid inputs and mentally simulating execution traces, eliminating the need for external sandboxes. Empirical evaluations based on LCB against state-of-the-art solutions and baselines demonstrate that the role of public test cases in driving correctness is much smaller than the community believes by matching the SOTA performance without using public test cases. While standard methods like CodeSIM exhibit an inflated overconfidence gap due to their reliance on trivial testing oracles, DryRUN maintains a much lower gap while maintaining comparable performance, proving the efficacy of execution-free self-correction. Ultimately, these findings challenge standard evaluation practices and raise the critical question of whether public test cases should be removed from coding benchmarks entirely to prevent biased, overconfident models in the field.


\section*{Data Availability}
The source code and datasets supporting the findings of this study are available at the following Zenodo repository: \href{https://zenodo.org/records/19348029?preview=1&token=eyJhbGciOiJIUzUxMiJ9.eyJpZCI6IjliYzkyMmNiLTgzYjktNDdkOC05NTk2LTgzOGU0MjdmOTcxZSIsImRhdGEiOnt9LCJyYW5kb20iOiJiZGUyYmMzN2FmNTc1NGY4ZGQyN2RhMzMwZjc4NDc3OCJ9.dkYW7wG70zc06ofDjBZqlO3z6Gmi05jBMtkRH4Ay6h-ZvmauMcmPwuNNyZnAtBugUGk94lafWgMOebVGomKE2A}{https://zenodo.org/records/19348029}.

\newpage
\bibliographystyle{ACM-Reference-Format}
\bibliography{references}

\end{document}